\documentclass[preprint]{aastex}
\usepackage{graphicx}

\newcommand{\xte}{XTE J1748$-$288}

\shorttitle{ASCA observations of XTE J1748$-$288}
\shortauthors{Kotani et al.}

\begin{document}

\title{ASCA OBSERVATIONS OF THE JET SOURCE XTE J1748$-$288}

\author{T. Kotani\altaffilmark{0},
N. Kawai\altaffilmark{1},
F. Nagase\altaffilmark{2},
M. Namiki\altaffilmark{1},
M. Sakano\altaffilmark{3},
T. Takeshima\altaffilmark{0},
Y. Ueda\altaffilmark{2},
K. Yamaoka\altaffilmark{2},
R. M. Hjellming\altaffilmark{4}}
\authoremail{kotani@milkyway.gsfc.nasa.gov,
nkawai@riken.go.jp, nagase@astro.isas.ac.jp,
namiki@riken.go.jp, sakano@cr.scphys.kyoto-u.ac.jp,
takeshim@olegacy.gsfc.nasa.gov, ueda@astro.isas.ac.jp,
yamaoka@astro.isas.ac.jp, rhjellmi@aoc.nrao.edu}

\altaffiltext{0}{Code 661, Laboratory for High-Energy Astrophysics, NASA/GSFC, Greenbelt, MD 20771, USA}
\altaffiltext{1}{Cosmic Radiation Laboratory, RIKEN, Saitama 351-0198, Japan}
\altaffiltext{2}{High-Energy Astrophysics Division, ISAS, Kanagawa 229-8510, Japan}
\altaffiltext{3}{Department of Physics, Kyoto University, Kyoto 606-8502, Japan}
\altaffiltext{4}{National Radio Astronomy Observatory, Socorro, NM 87801-0387, USA}

\begin{abstract}
XTE J1748$-$288 is a new X-ray transient with a one-sided radio jet.
It was observed with ASCA on 1998/09/06 and 1998/09/26, 100 days after
the onset of the radio-X-ray outburst.  The spectra were fitted with
an attenuated power-law model, and the 2--6-keV flux was
$4.6^{+1.0}_{-0.8} \times 10^{-11}$ erg s$^{-1}$ cm$^{-2}$ and
$2.2^{+0.8}_{-0.6}\times 10^{-12}$ erg s$^{-1}$ cm$^{-2}$ on 09/06 and
09/26, respectively.  The light curve showed that the steady
exponential decay with an $e$-folding time of 14 days lasted over 100
days and 4 orders of magnitude from the peak of the outburst.  The
celestial region including the source had been observed with ASCA on
1993/10/01 and 1994/09/22, years before the discovery.  In those
period, the flux was $\stackrel<\sim 10^{-13}$ erg s$^{-1}$ cm$^{-2}$,
below ASCA's detection limit.  The jet blob colliding to the
environmental matter was supposedly not the X-ray source, although the
emission mechanism has not been determined.  A possible detection of a
K line from highly ionized iron is discussed.
\end{abstract}

\keywords{stars: binaries  --- stars: individual (XTE J1748-288)}

\section{Introduction}
XTE J1748$-$288 showed an X-ray outburst on 1998/06/04 and detected by
the ASM/RXTE and BATSE/CGRO (Smith, Levine, \& Wood 1998; Harmon et
al.\ 1998).  The 2-10-keV X-ray flux reached to 600 mCrab on
1998/06/05 (Strohmayer et al.\ 1998), and decayed with an $e$-folding
time of $\sim 20$ days (ASM/RXTE Team, 1999; Sidoli et al.\ 1999).
The radio counterpart was located with the VLA at R.A. = 17h48m05s.06,
Decl.\ = $-28^\circ28'25".8$ (equinox 2000.0; uncertainty $0".6$;
Strohmayer et al.\ 1998).  The spectral index of the radio counterpart
was $0.2\sim0.6$ (Hjellming et al.\ 1998a; Rupen \& Hjellming 1998),
and the radio activity reached to the maximum of 350 mJy at 2.25 GHz
around 1998/06/16 (GBI Team 1999).  Rupen \& Hjellming discovered a
one-sided jet of 20 mas day$^{-1}$ in the VLA images, which
corresponds to a velocity of 0.93 $c$ assuming a distance of 8 kpc.
The time of the jet ejection was estimated to be around 1998/06/01,
extrapolating the proper motion (Hjellming et al.\ 1998b).  The radio
activity of the core lasted 3 months, suggesting a continuous ejection
of jet material.  Thus this source is considered to be a jet system,
probably similar to SS 433, which has a persistent jet.  Around
1998/08/09, the expansion of the leading edge slowed to a rate of
$\sim5$ mas day$^{-1}$ and the leading edge of the jet brightened
dramatically (Hjellming et al.\ 1998b).  Hjellming et al.\ suggested
that the ejected jet material has run into external gas and formed a
shock which is seen as a ``hot spot.''  On 1998/09/26, an ASCA TOO
observation of this source was performed (Kotani et al.\ 1999, 2000).
And the celestial region including the source had been observed with
ASCA on 1993/10/01, 1994/09/22, 1994/09/24, and 1998/09/20.  In this
paper we report the result of all of these observations of the
remarkable black hole candidate with a jet, which is possibly
interacting with the circumstellar matter as that of SS 433.

\section{Observation}
The log of the observations of XTE J1748$-$288 with ASCA is shown in
Table~\ref{kotani:tbl:log}.  In observation no.\ 4, \xte\ was near the
center of the FOV and data from both the GIS and SIS were obtained.
Other observations were performed as parts of the
Galactic-plane-survey program (Koyama et al.\ 1997) or the
Galactic-center-region observations (Maeda et al.\ 1999; Sakano et
al.\ 1999), and yet no SIS data were obtained because \xte\ was out of
SIS' FOV\@.  Observation 2B were performed soon after 2A, and they are
regarded as one data set hereafter.  A simultaneous observation with
RXTE was performed during observation no.~4 (Revnivtsev, Trudolyubov, \&
Borozdin 2000).

In all observations, the PH mode with the nominal bit assignment was
used for the GIS (Tanaka, Inoue, \& Holt 1994).  In observation no.\ 4, the
1-CCD FAINT mode was used for the SIS\@.  All data were processed with
the standard event-selection criteria and the data-reduction method (ASCA
Guest Observer Facility 1999), unless specified in text.

\placetable{kotani:tbl:log}

\section{Data Analysis}
\subsection{Image}
The GIS image taken in no.\ 4, when \xte\ was near the center of the
FOV, is shown in
Fig.~\ref{kotani:fig:gisimage}.  
Because \xte\ was rather faint at that time, other sources are
recognizable by eye in the figure.  In addition to XTE J1748$-$288, 1E
1743.1$-$2843, which is the brightest source in the FOV, Sgr B2, and
an unidentified source at R.A. = 17h47m$04\pm15$s, Decl.\ =
$-28^\circ53'\pm1'$ are seen in the image.  The extended source in the
background of these sources is the galactic ridge emission.  Sgr B2,
1E 1743.1$-$2843, and the galactic ridge emission can be sources of
background photons.  Especially, Sgr B2 and the galactic ridge are
significant iron-line emitters (Murakami et al.\ 2000), and their
photons should be removed carefully from the spectrum of XTE
J1748$-$288.

\placefigure{kotani:fig:gisimage}

\subsection{Light Curve}
Temporal activity of \xte\ was estimated from the GIS data.  We
collected photons within $6'$ of XTE J1748$-$288 for no.~3, when the
source was rather bright, and $2'$ for other data sets.  To estimate
the background spectrum of each observation, photons were collected
from the annular region around the source for each data set.  The
thickness of the annulus was $6'-10'$ for no.~4, and $6'-8'$ for
others.  Assuming that the background count rate is proportional to
the sampling area, we estimated the contribution by the background
component in the source region for each data set from the photons in
the annular region.

The source was significantly detected from no.~3 and~4, but not from 1
or 2.  The 2--6-keV flux was determined to be $4.6^{+1.0}_{-0.8}
\times 10^{-11}$ erg s$^{-1}$ cm$^{-2}$ and $2.2^{+0.8}_{-0.6}\times
10^{-12}$ for no.~3 and~4, respectively.  The 90 \% upper limit of
2--6-keV flux was $9.3 \times 10^{-13}$ erg s$^{-1}$ cm$^{-2}$ and
$4.9 \times 10^{-13}$, for no.~1 and~2, respectively.  The light curve
is shown in Fig.~\ref{kotani:fig:lc} together with the ASM/RXTE and
BeppoSAX data.

\placefigure{kotani:fig:lc}

\subsection{Spectrum}
In no.~3, \xte\ was rather bright, and that made the
background-component estimation less difficult.  The background
spectrum used to plot the light curve was also used for spectral fits.
The GIS spectrum of no.~3 is shown in Fig.~\ref{kotani:fig:gisspec3}.
Because the flux level of \xte\ was almost marginal to detect on
1998/09/26, no.~4, the background component should be subtracted very
carefully.  The $6'-8'$ annular region used for the background
estimation to plot the light curve is not sufficiently accurate, since
the extended background source has a structure finer than a few arcmin
as seen in Fig.~\ref{kotani:fig:gisimage}.  To minimize the
uncertainty due to the selection of a background region, we obtained a
background spectrum from the same sky region as the source region.  We
collected photons within $2'$ of \xte\ from no.~1 and~2, when the
source was inactive and fainter than the detection limit, and used
them as the background component.  The intensity of the non X-ray
background depends on the position in the FOV, and might cause a
systematic error in the background estimation (Makishima et al.\
1996).  However, the position dependence is significant in the soft
band below 1 keV, and did not affect the discussion following.  After
correction of the vignetting effect of the GIS, the resultant
background was subtracted from the source spectrum of no.~4.  The GIS
spectrum of no.~4 are shown in Fig.~\ref{kotani:fig:gisspec4}.  Even
after these careful treatment, the background estimation might have a
systematic uncertainity.  The negative data points below 2 keV might
be due to over-estimation of the background component.

We tried spectral fits of the GIS data with several spectral models.
The statistics did not allow us to determine emission mechanism
uniquely.  The spectra are well expressed by an attenuated power-law
model, an attenuated thin-thermal plasma emission (Mewe et al.\ 1985),
and an attenuated blackbody model.  The best-fit parameters of each
model are shown in Table~\ref{kotani:tbl:bestfit}.

\placetable{kotani:tbl:bestfit}

Although the goodness of the fits was already satisfactory, inclusion
of a Gaussian line at iron K energy into the power-law model
improved the fit.  The parameters of the Gaussian model are shown in
Table~\ref{kotani:tbl:gaussian}.  Even after the inclusion of a
Gaussian line at 6.9 keV into the fit model, the residuals of no.~3
still showed a feature which can be fitted with a line model at 5.9
keV (Kotani et al.\ 2000).  

\placetable{kotani:tbl:gaussian}

\xte\ was in the FOV of the SIS only in no.~4, but not in other
observations.  That made the background subtraction of the SIS
difficult.  Since the FOV of the SIS is not so large as that of the
GIS, photons farther than $4.5'$ from the source in the CCD chip were
used for the background estimation.  Because the effective exposure
time of no.~4 is only 20 ks, the statistics of the resultant
background spectrum was not good.  Spectral fits of the SIS were found
to be consistent with those of the GIS\@.

\section{Discussion}
\subsection{Temporal behavior}
In the light curve, the ASCA no.~3 (MJD = 51062) is on the
extrapolation line from ASM and BeppoSAX data.  It is remarkable that
the exponential decay with an $e$-folding time of $12 \sim 16$ days
lasted till no.~3 from the peak of the outburst, over 100 days and~4
orders of magnitude.  This is one of the best examples of the steady
exponential decay following an outburst shown by a black-hole
candidate.  On the other hand, the data point of no.~4 (MJD = 51082)
is below the extrapolation line by one order of magnitude.  There
might be a transition from the exponential-decay phase to another
phase, say, off-state phase, in the period between the two ASCA
observations.  In decaying phase of black-hole candidates, it is usual
to observe a hump in the light curve or a relaxation of the time
scale of decay.  The sudden drop in the light curve of \xte\ is
distinctive from the behaviors of such black-hole candidates.

The flux of no.~4 was inconsistent with the value of 10
mCrab obtained with RXTE (Revnivtsev et al.\ 2000).  Spectral analysis
with the RXTE is supposed to be difficult when the source is faint
because of the many background sources in RXTE's FOV\@.  As
Revnivtsev et al.\ suggested in their paper, the RXTE flux might be
contaminated by nearby sources.

It was fortunate that the source had been monitored with ASCA five
years before the discovery.  Although the data of no.~1 and~2 provided
only upper limits, they constrain the flux level of black-hole
candidates before onset of activity.  The mass-overflow instability
model (e.g., Hameury et al.\ 1986) predicts a certain hard X-ray
luminosity to trigger an outburst even in quiescent phase.  Assuming a
power-law spectrum with parameters same as no.~3 in
Table~\ref{kotani:tbl:bestfit} and a distance of 8 kpc, the upper
limit of luminosity above 7 keV in no.\ 1 and~2 would be $6.1 \times
10^{33}$ erg s$^{-1}$ and $3.2 \times 10^{33}$ erg s$^{-1}$,
respectively.  Such a low luminosity level of hard X ray is difficult
to trigger an outburst via the mass-overflow instability, unless the
companion star is lighter than 0.4 M$_\odot$, according to the
estimation by Mineshige et al.\ (1992).  That favors the
disk-instability scenario (e.g.\ Osaki 1996) rather than the
mass-overflow-instability as the cause of the outburst of \xte.

\subsection{Emission mechanism}
It is a difficult question to answer whether the iron line detected in
no.~3 and~4 was intrinsic to XTE J1748$-$288, or that it was
originated in the background Galactic ridge emission, even after the
careful background subtraction described above.  The negative flux
below 2 keV of no.~4 suggests that the background component was
slightly over-estimated, and thus the excess at the iron K energy was
a significant real structure.  We tried several background-subtraction
methods other than given above, and recognized the iron line in all
cases.  On the other hand, the absence of any iron line in the
BeppoSAX spectrum of XTE J1748$-$288 on 1998/08/26 casts a serious
doubt, though the source was at off-center of the MECS, where the
energy resolution is not the best (Sidoli et al.\ 1999).  The
possibility that the iron line was not related to \xte\ could not be
rejected at this stage.  The origin of the iron line shall be
determined by a future observation with a mission of higher
sensitivity and spatial resolution.

If the iron line was of \xte\ origin, the emission mechanism may be
thin-thermal or fluorescent.  A hot jet is of special interest among
possible sources such as an illuminated accretion disk, optically thin
advection disk, and accretion column.  Physical parameters of the
system may be determined from the iron line of a jet.  If the line was
Doppler-shifted Fe {\sc xxv} K$\alpha$, the Doppler-shift parameter
$z$ was determined as $(1-z)^{-1} = 1.012^{+0.025}_{-0.027}$.  From
the proper motion $\mu = 0.93\times (D/\mbox{8 kpc})$ (Rupen \&
Hjellming 1998), the Lorentz factor $\gamma$ and inclination $j$ were
estimated as $1 + 0.41 \times ( D/ \mbox{8 kpc} )^2 < \gamma < 1+ 0.44
\times ( D/ \mbox{8 kpc} )^2$ and $1.9 < \tan j < 2.2$, respectively.
Considering that $D$ may be larger than 8 kpc, we constrained $\beta$
and $j$ as $\beta > 0.71$ and $j<72^\circ$.  Kotani et al.\ (1996)
modeled SS 433's jet as a conical plasma flow cooling by radiation and
expansion to explain ASCA data.  We adopted the code for \xte, and
estimated the mass outflow rate.  In the numerical calculation, the
velocity and the inclination of the jet, and the distance to the
source were set to $\beta=0.73$ $j=64^\circ$, and $D = 8$ kpc,
respectively.  Other input parameters were set to the same values as
those of SS 433 (Kotani 1998) Assuming that the X-ray source is a hot
jet, we estimated the mass outflow rate to be the order of $10^{-5}$
M$_\odot$ y$^{-1}$ and $10^{-6}$ M$_\odot$ y$^{-1}$ on no.~3 and~4,
respectively.  The mass outflow rate of no.~3 is comparable to that of
SS~433 (Kotani 1998).

The ejected blob of \xte\ had been observed to brighten and decelerate
to 0.23 $\times (D/\mbox{8 kpc})$ $c$ (Rupen \& Hjellming 1998),
probably crashing to circumstellar matter.  The shock temperature of
the blob decelerated from 0.93 $c$ to 0.23 $c$ would be $\sim 100$
MeV, if the blob consists of baryonic plasma.  Such plasma might emit
X ray via bremsstrahlung or synchrotron process.  The spectrum would
not have a line due to the high temperature.  However, the temporal
behavior of the X-ray luminosity of such plasma would be different
from the observations.  Plasma cooling via radiation or expansion will
not show an exponential decay, but a power-law light curve.  Thus the
colliding blob was supposedly not the dominant X-ray source, i.e., the
X rays from the blob had decayed quickly before no.~3, or it was
faint.  If plasma with a temperature of 100 MeV was cooled within 10
days due to radiation, the density must be larger than $10^{11}$
cm$^{-3}$, a quite large value for a jet blob at the end.  If the
colliding blob was faint, the upper limit of the flux,
$2.2^{+0.8}_{-0.6}\times 10^{-12}$, gives an upper limit of the
density of the blob, $\stackrel < \sim 10^3$ cm$^{-3}$, assuming a
blob dimension of $10^{16}$ cm.  That gives an upper limit of the
ejected mass of $10^{28}$ g, or $10^{-6}$ M$_\odot$.

As conclusion, the colliding jet blob was not the dominant X-ray
source in observation no.~3, whether the line in
Table~\ref{kotani:tbl:gaussian} was related with \xte\ or not.  The
featureless spectrum may be explained in terms of hard power-law like
emission seen in black hole candidates in low state.  As for no.~4,
the possibility of the emission from the colliding blob could not be
rejected based on the temporal behavior.  The X rays in no.~4 might be
emitted from or contaminated by the blob.

\section{Summary}
The new Galactic jet source \xte\ was observed with ASCA four times,
twice in the quiescent phase before the outburst, and twice 100 days
after the onset of the outburst.  The flux was below $10^{-13}$ erg
s$^{-1}$ cm$^{-2}$ in the quiescent phase, and $4.6^{+1.0}_{-0.8}
\times 10^{-11}$ erg s$^{-1}$ cm$^{-2}$ and $2.2^{+0.8}_{-0.6}\times
10^{-12}$ on 09/06 and 09/26, respectively.  The light curve decayed
exponentially with an $e$-folding time of $12-16$ days, over 100 days
and 4 orders of magnitude.  The temporal behavior suggested that the
colliding jet blob was not the dominant X-ray source, at least on
1998/09/06.  There was an indication of an iron line in the spectra
after the careful background subtraction.  A hot jet, an accretion
column, or an accretion disk may account the X-ray emission.  Assuming
that the jet being ejected was the X-ray source, we constrained the
velocity and the inclination of the jet as $\beta > 0.71$ and $j <
72^\circ$, respectively.  The upper limit of the flux would be
constrained the ejected matter less than $10^{28}$ g.

\vspace{1cm}

The ASCA data were obtained in a TOO observation by courtesy of the
ASCA Team, and in the Galactic plane survey program.  The radio data
were provided via the public archive of the Green Bank Interferometer,
which is a facility of the National Science Foundation operated by the
NRAO in support of NASA High Energy Astrophysics programs.  TK is
supported by the research associateship program of the National
Research Council.

\clearpage
\noindent
{\large\bf References}
\begin{description}
\item ASCA Guest Observer Facility, 1999, {\tt
http://heasarc.gsfc.nasa.gov/docs/asca/abc/abc.html} 
\item ASM/RXTE Team 1999, quick-look results provided via {\tt http://space.mit.edu/XTE/ASM\_{}lc.html}
\item GBI Team 1999, quick-look results provided via {\tt http://www.nrao.edu/\~{}rhjellmi/gbint/plgbi.html}

\item Hameury, J-M., King, A.R., \& Lasota, J-P.\ 1986, \aa, 162, 71

\item Harmon, B. A., McCollough, M. L., Wilson, C. A., Zhang, S. N., \&
Paciesas, W. S. 1998, IAUC, 6933
\item Hjellming, R. M., Rupen, M. P., Ghigo, F., Fender, R. P., \& Stappers,
B. W. 1998a, IAUC, 6937
\item Hjellming, R. M., Rupen, M. P., Mioduszewski, A. J.,
                    Smith, D. A., Harmon, B. A., Waltman, E. B.,
                    Ghigo, F. D., \& Pooley, G. G. 1998b, \aaps, 193, 10308
\item Kotani, T., Kawai, N., Matsuoka, M., \& Brinkmann, W. 1996, \pasj, 48, 619
\item Kotani, T. 1998, Doctoral Thesis, University of Tokyo
\item 12. Kotani, T., Band, D., Cherepashchuk, A. M., Hjellming, R. M., Kawai, N.,
 Matsuoka, M., Namiki, M., Oka, T., Shirasaki, Y., Tsutsumi, T. 1999,
 Astronomische Nachrichten 320, 335

\item Kotani, T., Kawai, N., Nagase, F., Namiki, M., Sakano, M.,
Takeshima, T., Ueda, Y., \& Matsuoka, M. 2000, Astrophys.\ Lett.\
Communications, submitted
\item {Koyama}, K., {Yamauchi}, S., {Sugizaki}, M., \& The ASCA
Galactic Plane Survey Team, 1997, in All-Sky X-Ray Observations in the
Next Decade, M. Matsuoka \& N. Kawai (Saitama, RIKEN), 45

\item {Maeda}, Y., {Koyama}, K., {Imanishi}, K., {Murakami}, H., 
        {Nishiuchi}, M., {Sakano}, M., {Tsujimoto}, M., {Yokogawa}, J., \&
        {Yamauchi}, S. 1999, Astronomische Nachrichten, 320, 177 

\item   {Makishima}, K., {Tashiro}, M., {Ebisawa}, K., {Ezawa}, H., 
        {Fukazawa}, Y., {Gunji}, S., {Hirayama}, M., {Idesawa}, E., 
        {Ikebe}, Y., {Ishida}, M., {Ishisaki}, Y., {Iyomoto}, N., et
al.\ 1996, \pasj, 48, 171

\item Mewe, R., Gronenschild, E. H. B. M., \& van den Oord,
G. H. J. 1985, A\&AS, 62, 197

\item Mineshige, S., Ebisawa, K., Takizawa, M., Tanaka, Y., Hayashida,
K., Kitamoto, S., Miyamoto, S., \& Terada, K. 1992, \pasj, 44, 117 

\item    Murakami, H., Koyama, K., Sakano, M., Tsujimoto, M., \&
Maeda, Y. 2000, \apj, in press

\item Osaki, Y. 1996, \pasp, 108, 39

\item Revnivtsev, M. G., Trudolyubov, S. P. and Borozdin, K. N. 2000, \mnras, 
312, 151

\item Rupen, M. P., \& Hjellming, R. M. 1998, IAUC, 6938

\item Sakano, M., Koyama, 
K., Yokogawa, J., Murakami, H., Nishiuchi, M., Maeda, Y., \& Yamauchi, S. 
1999, Astronomische Nachrichten, 320, 330

\item Sidoli, L., Mereghetti, S., Israel, G. L., Chiappetti, L,
Treves, A., \& Orlandini, M. 1999, \apj, 525, 215
\item Smith, D.A., Levine, A., \& Wood, A. 1998, IAUC, 6932
\item Strohmayer, T., Marshall, F. E., Hjellming, R.M., \& Rupen
M. P. 1998, IAUC, 6934
\item Tanaka, Y., Inoue, H., \& Holt, S. S. 1994, \pasj, 46, L37
\end{description}


\figcaption[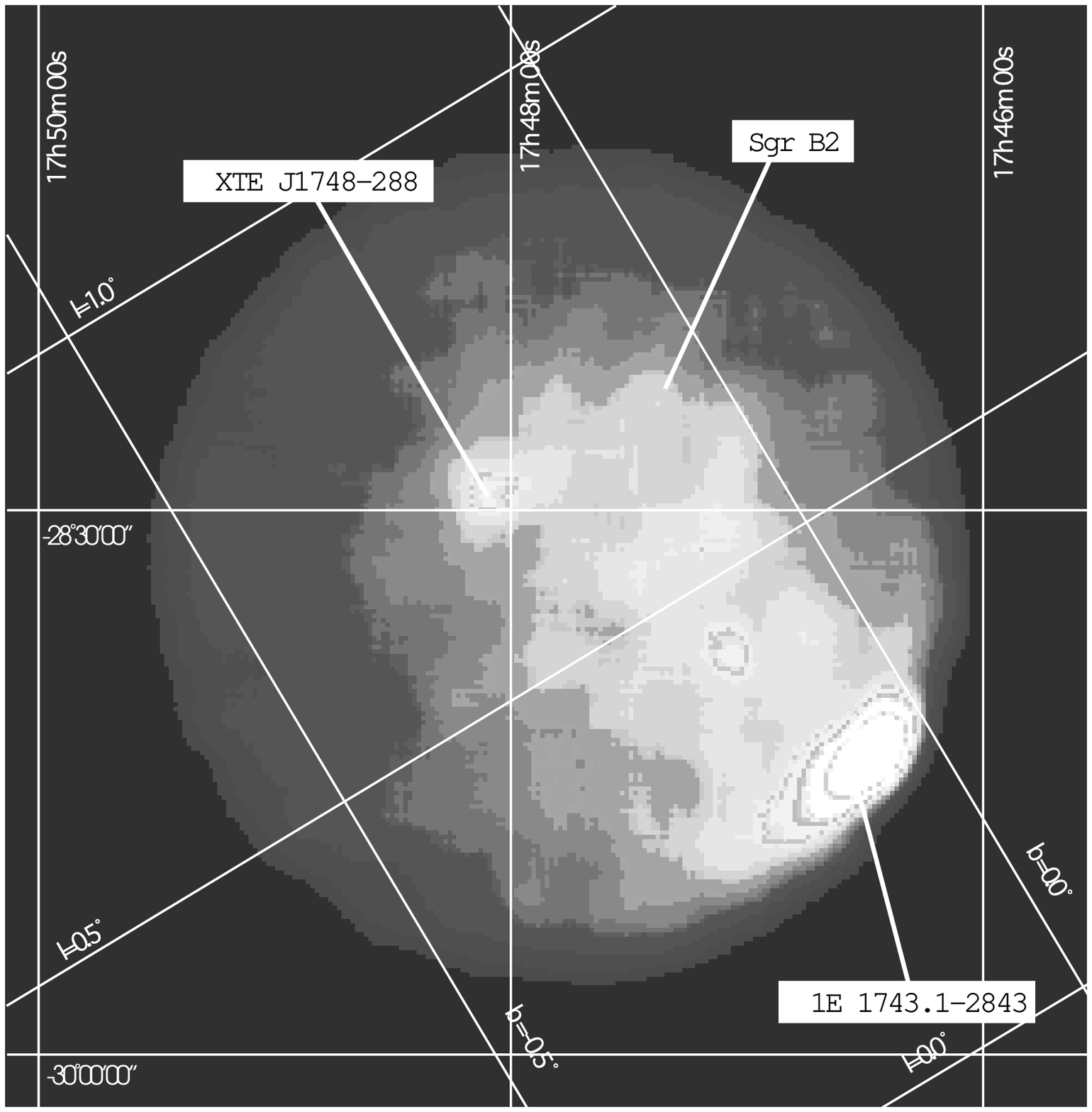] {GIS image taken in observation no.~4.  The energy
band of 2.9--5.9 keV was used.  Data of both GIS2 and GIS3 were
combined and convolved with a 2-dimensional Gaussian ($\sigma=3$
pixels).  The X rays from the calibration sources were removed.  XTE
J1748$-$288, the nearby sources 1E 1743.1$-$2843 and Sgr B2 are
indicated.  There is a point source at R.A. = 17h47m04s, Decl.\ =
$-28^\circ53'$.
\label{kotani:fig:gisimage}}

\figcaption[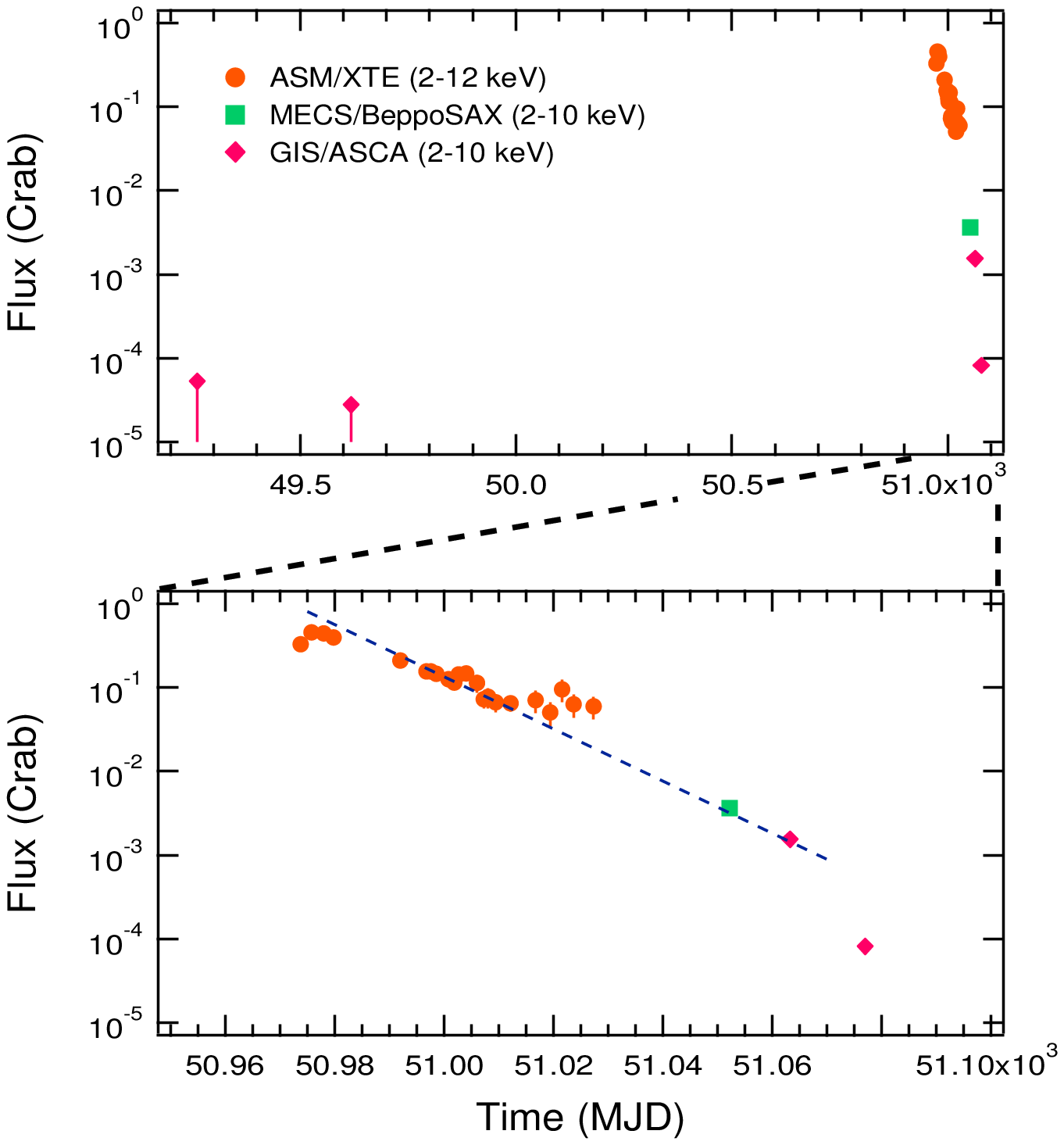] {Light curve of \xte.  The data of the ASM/RXTE
(ASM/RXTE Team 1999), the MECS/BeppoSAX (Sidoli et al.\ 1999), and
ASCA are plotted.  The dashed line is an exponential decay with an
$e$-folding time of 14 days.  The convergence of the ASM data to $\sim
0.07$ Crab is artificial due to the data selection threshold applied
here.
\label{kotani:fig:lc}}

\figcaption[fig3.ps] {GIS spectrum of XTE J1748$-$288 of no.\ 3.  Sum
of the two sensors.  The histogram is the best-fit power-law model.
The data were rebined for display.
\label{kotani:fig:gisspec3}}

\figcaption[fig4.ps] {GIS spectrum of XTE J1748$-$288 of no.\ 4.  Sum
of the two sensors.  The histogram is the best-fit power-law model.
The data points below 2 keV are negative probably due to
over-estimation of the background flux, while the points around iron K
energy show an excess.  The data were rebined for display.
\label{kotani:fig:gisspec4}}



\clearpage

\begin{table}
\caption{Observation log}\label{kotani:tbl:log}
\begin{tabular}{ccccl}
\tableline\tableline
No.	&Start (MJD)	&End (MJD)	&Effective Time	&Remark\\
	&	&	& (ks)	&\\
\tableline
1	&1993/10/01 21:20 (49261.89)	&1993/10/02 10:20 (49262.43)
&15	&GIS only\\
2A	&1994/09/22 03:50 (49617.16)	&1994/09/23 12:10 (49618.51)
&58	&GIS only\\
2B	&1994/09/24 02:00 (49619.08)	&1994/09/24 14:40 (49619.08)
&20	&GIS only\\
3	&1998/09/06 09:10 (51062.38)	&1998/09/06 17:50 (51062.74)
&5 	&GIS only\\
4	&1998/09/26 03:30 (51082.15)	&1998/09/26 16:30 (51082.69)
&20	&GIS+SIS\\
\tableline
\end{tabular}
\end{table}

\clearpage

\begin{table}
\caption{Best-fit parameters}\label{kotani:tbl:bestfit}
\begin{tabular}{cccc} \tableline\tableline
\multicolumn{4}{c}{Power Law}\\
\tableline
No.	&$N\rm_H$	&$\Gamma$	&red. $\chi^2$ (d.o.f.)\\
	&(cm$^{-2}$)	&	&\\
\tableline
3	&$6.0^{+1.3}_{-1.1}\times10^{22}$	&$1.56^{+0.34}_{-0.31}$	&0.94 (180)\\
4	&$18.5^{+10.9}_{-7.8}\times10^{22}$	&$8.1^{+8.3}_{-3.2}$	&1.05 (187)\\
\tableline\tableline
\multicolumn{4}{c}{Blackbody}\\
\tableline
No.	&$N\rm_H$	&$kT$	&red. $\chi^2$ (d.o.f.)\\
	&(cm$^{-2}$)	&(keV)	&\\
\tableline
3	&$2.86^{+0.80}_{-0.68}\times10^{22}$	&$1.72^{+0.21}_{-0.18}$	&1.00     (180)\\
4	&$13.3^{+21.4}_{-6.1}\times10^{22}$	&$0.39^{+0.30}_{-0.19}$	&1.05 (187)\\
\tableline\tableline
\multicolumn{4}{c}{Thin-Thermal Plasma}\\
\tableline
No.	&$N\rm_H$	&$kT$	&red. $\chi^2$ (d.o.f.)\\
	&(cm$^{-2}$)	&(keV)	&\\
\tableline
3	&$5.65^{+0.89}_{-0.91}\times10^{22}$	&$>11.2$	&0.94 (180)\\
4	&$14.5^{+22.6}_{-5.7}\times10^{22}$	&$0.67^{+0.67}_{-0.45}$	&1.05 (187)\\
\tableline\tableline
\end{tabular}
\tablecomments{For flux, see  the text.}
\end{table}

\clearpage

\begin{table}
\caption{Best-fit Gaussian parameters}\label{kotani:tbl:gaussian}
\begin{tabular}{cccccc}
\tableline
No.	&Centroid Energy	&$\sigma$	&Flux	&d.o.f. ($\Delta$d.o.f.)	&$\Delta\chi^2$\\
	&(keV)	&(keV)	&(ph cm$^{-2}$ s$^{-1}$)\\
\tableline
3	&$6.73^{+0.23}_{-0.12}$&$<0.28$&$2.7^{+2.0}_{-2.0}\times10^{-4}$	&177 ($-3$)	&$-4.51$\\
4	&$6.92^{+0.16}_{-1.26}$	&$<0.66$	&$2.5^{+2.0}_{-1.9}\times10^{-5}$	&184 ($-3$)	&$-5.35$\\
\tableline
\end{tabular}
\end{table}







\end{document}